\documentclass[showpacs,preprintnumbers,amsmath,amssymb,aps,prl,twocolumn]{revtex4-1}

\usepackage{amsmath,amssymb}
\usepackage{graphicx}
\usepackage{dcolumn}
\usepackage{bm}
\usepackage{color}
\usepackage{times}
\usepackage{wasysym}
\usepackage{times}
\usepackage{multirow}
\usepackage{verbatim}

\renewcommand{\eqref}[1]{(\ref{eq:#1})}

\newcommand{\figref}[1]{Fig.~\ref{fig:#1}}
\newcommand{\Figref}[1]{Figure~\ref{fig:#1}}

\begin{document}

\title{Topology-optimized Dual-Polarization Dirac Cones}

\author{Zin Lin$^{1}$}
\thanks{These authors contributed equally to this work.}

\author{Lysander Christakis$^{2}$}
\thanks{These authors contributed equally to this work.}
\author{Yang Li$^{1}$}
\thanks{These authors contributed equally to this work.}

\author{Eric Mazur$^1$}
\author{Alejandro W. Rodriguez$^{3}$}
\author{Marko Lon\v{c}ar$^1$}
\email{loncar@seas.harvard.edu}
\affiliation{$^1$John A. Paulson School of Engineering and Applied Sciences, Harvard University, Cambridge, MA 02138}
\affiliation{$^2$Department of Physics, Yale University, New Haven, CT 06511}
\affiliation{$^3$Department of Electrical Engineering, Princeton University, Princeton, NJ, 08544}

\date{\today}

\begin{abstract}
We apply a large-scale computational technique, known as topology optimization, to the inverse design of photonic Dirac cones. In particular, we report on a variety of photonic crystal geometries, realizable in simple isotropic dielectric materials, which exhibit dual-polarization Dirac cones. We present photonic crystals of different symmetry types, such as four-fold and six-fold rotational symmetries, with Dirac cones at different points within the Brillouin zone. The demonstrated and related optimization techniques open new avenues to band-structure engineering and manipulating the propagation of light in periodic media, with possible applications to exotic optical phenomena such as effective zero-index media and topological photonics. 

\end{abstract}

\maketitle

Dirac cones (DC), or conical dispersions, have received broad
attention due to their special properties affecting light transport in
photonic systems, such as effective zero-index
behavior~\cite{Chan2011,moitra2013realization,li2015chip,Hajian2016,He2016,Engheta2017,Kita2017}, exceptional
points~\cite{zhen2015,lin2016inverse}, photonic
\emph{Zitterbewegung}~\cite{zhang2008observing}, and topologically
protected states~\cite{lu2014topological,khanikaev2013photonic}. So
far, DCs have been primarily studied in simple geometries based on
circular pillars or air holes on a periodic
lattice~\cite{Chan2011,moitra2013realization,li2015chip,Vulis2016chip,He2016,Kita2017}. One
exception is our previous work~\cite{lin2016inverse}, which exploited
topology optimization (TO) techniques to demonstrate higher-order DCs
(precursors to exceptional points ~\cite{zhen2015,lin2016inverse}) in complex structures.
TO, which was first proposed more than a decade ago~\cite{sigmund2003}, employs gradient-based algorithms to efficiently handle a very
large design space, considering every pixel or voxel as a degree of
freedom (DOF) in an extensive 2D or 3D computational
domain~\cite{Jensen11}. Recently, inverse-designed materials based on TO have been utilized to improve the
performance of optical devices such as mode splitters,
de-multiplexers, and wavelength
converters~\cite{Jensen11,liang2013formulation,piggott2015inverse,lu2013nanophotonic,shen2015integrated,lin2016cavity}. 
Here, we apply TO toward the design of unprecedented dispersion
features in photonic crystals (PhC), namely, two overlapping DCs with
dual polarizations (DPDC): one having transverse magnetic polarization
($\mathbf{H}\cdot \mathbf{\hat{z}}=0$) and the other having transverse
electric polarization ($\mathbf{E}\cdot \mathbf{\hat{z}}=0$). We show
that if designed at the $\mathbf{\Gamma}$ point of the Brillouin zone, such
PhCs exhibit effective zero-index behavior. Furthermore, we
demonstrate DPDCs at the $\mathbf{K}$ point of a hexagonal PhC, which
has implications for all-dielectric topological photonics~\cite{khanikaev2013photonic, pti3d16}.

Recent years have witnessed an exciting quest for exotic composite
materials along with unusual states of matter involving enhanced optical,
mechanical, and quantum
properties~\cite{wang2016lightweight,drozdov2015conventional,qin2017mechanics,wong2016lasing}. However, there has been comparatively less effort devoted to
discovering unconventional structures that can enhance the
functionality of ordinary materials, such as
ubiquitous low-loss isotropic dielectrics. Our work represents an
effort to leverage the capabilities of established but under-utilized
inverse design tools to uncover increased functionalities for simple
dielectrics.

\begin{figure*}[ht!]
\centering
\includegraphics[width=0.95\textwidth]{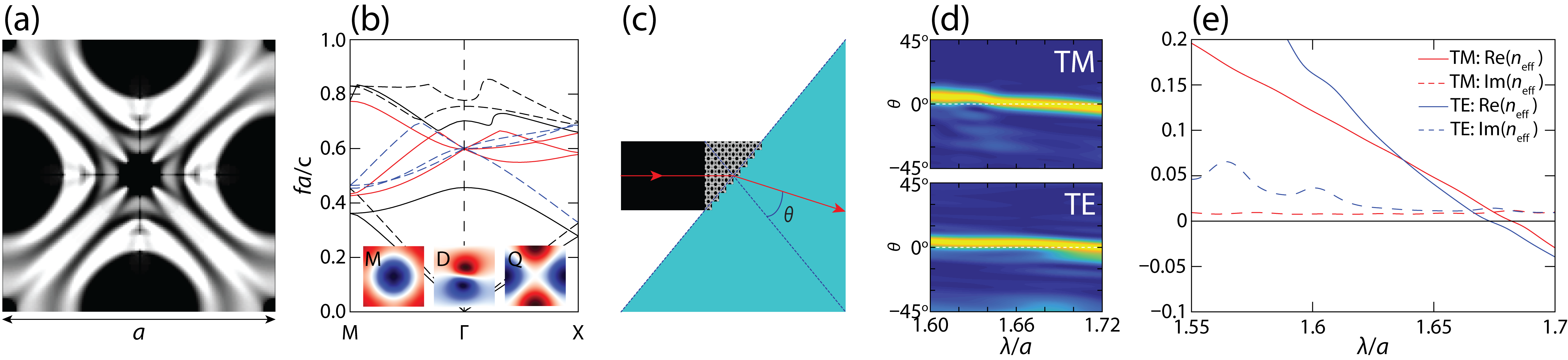} \caption{(a) Topology-optimized unit cell. Black (white) regions have relative permittivity $\epsilon_{\rm r} \approx 5.5$ ($\epsilon_{\rm r}=1$). Gray regions with intermediate permittivities are also seen. Note that the structure obeys $C_{4v}$ symmetry. (b) The band structure reveals two overlapping Dirac cones, one for TM polarization (solid lines) and the other for TE polarization (dashed lines). Transverse magnetic Dirac bands (dark red lines) are formed by the degeneracy of one monopolar (M) and two dipolar (D) modes manifested by the $E_z$ component whereas transverse electric Dirac bands (light blue lines) are formed by the degeneracy of two dipolar (D) and one quadrupolar (Q) modes manifested by the $H_z$ component (see figure inset). (c) Configuration of the prism test designed to illustrate effective zero index behavior for designs with dual polarization Dirac cones (DPDC). (d) FDTD analysis of the DPDC structures and their farfield patterns through the prism test show orthogonally emerging beams at the prism facet ($\theta$ = 0), validating the effective zero index behavior for both TM- and TE-polarized waves incident on the non-binary design. Also shown are the retrieved TM and TE effective indices for the non-binary design (e).\label{fig:fig1}}
\end{figure*}

\begin{figure*}[ht!]
\centering
\includegraphics[width=0.95\textwidth]{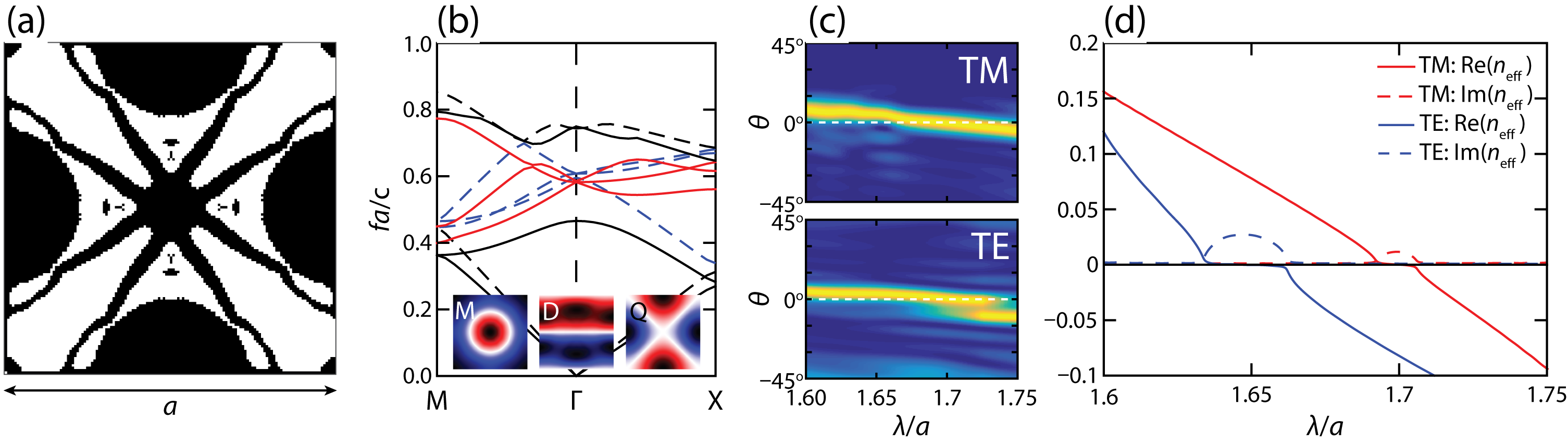}
\caption{(a) Binary regularized version of DPDC PhC unit cell, with the corresponding band structure (b) showing TM (solid lines) and TE (dashed lines) Dirac cones. FDTD analysis and the farfield patterns through the prism test show orthogonally emerging beams at the prism facet $(\theta=0)$, validating the effective zero index behavior for both TM- and TE-polarized incident waves. (d) Also shown are the TM and TE effective indices retrieved from scattering coefficients. \label{fig:fig2}}
\end{figure*}

{\it Dual-polarization Dirac Cones.---} Power emitted by a
time-harmonic current source $\mathbf{J} \sim e^{i \omega t}$,
proportional to the local density of states (LDOS), offers a
convenient optimization framework for designing eigenmodes at a given
frequency $\omega$. This follows the well-known principle that emitted
power, $f(\mathbf{E},\mathbf{J},\omega;\epsilon)= -\mathrm{Re} \Big[
  \int \mathbf{J}^*\cdot\mathbf{E}~d\mathbf{r} \Big]$, is maximized
when the source couples to a long-lived
resonance~\cite{Liang13}. Here, the electric field $\mathbf{E}$ is
simply the solution of the steady-state Maxwell equation, $\nabla
\times {1 \over \mu} \nabla \times \mathbf{E} - \omega^2
\epsilon(\mathbf{r}) \mathbf{E} = i \omega \mathbf{J}$. The goal of TO
is to discover the dielectric profile $\epsilon(\mathbf{r})$ that
maximizes $f$ for any given $\mathbf{J}$ and $\omega$. In what
follows, we judiciously choose $\mathbf{J(\mathbf{r})}$ and the
symmetries of the unit cell to construct PhCs with a variety of
intriguing spectral features (see supplemental material for details). In particular, we apply TO to design \emph{six accidentally
  degenerate} modes of monopolar (M), dipolar (D) and quadrupolar (Q)
profiles (inset of \figref{fig1}) that transform according to the $A$ and
$E$ irreducible representations of the $C_{4v}$ point group, and which,
in turn, give rise to conical dispersions in the vicinity of their
degeneracy~\cite{Sakoda12,mei2012first}. We emphasize that designing
such a \emph{six-fold} degeneracy poses a significant challenge for
conventional design and even for sophisticated, heuristic optimization
algorithms such as particle swarms, simulated annealing, or genetic
algorithms~\cite{anneal,genetic,swarm}, but can be susceptible to efficient gradient-based
TO techniques, in combination with
a proper problem formulation.

\Figref{fig1} shows a topology-optimized PhC unit cell and its
associated band structure, which exhibits two overlapping DCs at
the $\Gamma$ point, one with transverse electric (TE) and the other with
transverse magnetic (TM) polarization. Note that the DC in our design is the so-called generalized DC typically characterized by the presence of an extra flat band~\cite{Dóra2011}. Within the DPDC, TM Dirac bands
are formed by the degeneracy of one monopolar (M) and two dipolar (D)
modes, whereas TE Dirac bands are formed by the degeneracy of two
dipolar (D) and one quadrupolar (Q) modes. The optimized structure
consists of high dielectric regions ($\epsilon_{\rm r}=5.5$), typical of common materials such as silicon nitride or titania, in a
background of air ($\epsilon_{\rm r}=1$). Intermediate permittivities,
$\epsilon_{\rm r} \in (1,5.5)$, are also seen as a result of
fine-tuning the necessary modal frequencies to ensure a perfect
overlap. The resulting \emph{gray-scale} PhC has altogether
\emph{six} DPDC modes whose frequencies are degenerate to within
$0.1\%$, an accuracy limited only by numerical resolution. Here, we note that in a few initial rounds of optimization, we deliberately optimize for a TM quadrupole at $f_{\rm TMQ} \sim 0.75f_{\rm D}$ and a TE monopole at $f_{\rm TEM} \sim 1.25f_{\rm D}$ \textit{in addition} to the six degenerate modes forming the DC. We find that such a procedure for ``mode separation" helps ensure well-isolated conical dispersions.

\begin{figure*}[ht!]
\centering
\includegraphics[width=0.95\textwidth]{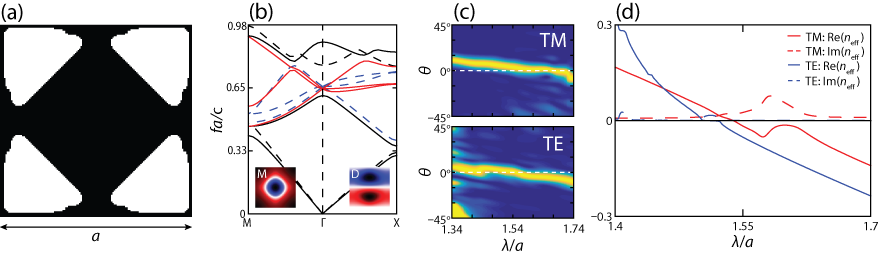}
\caption{(a) Binary regularized DPDC PhC with fabrication-friendly features. (b) The corresponding band structure shows two overlapping TM (solid lines) and TE (dashed lines) Dirac cones. (c) FDTD analysis and the farfield patterns through the prism test [\figref{fig3}(a)] show orthogonally emerging beams at the prism facet $(\theta=0)$, validating the effective zero index behavior for both TM- and TE-polarized incident waves. (d) Also shown are the TM and TE retrieved effective indices. \label{fig:fig3}}
\end{figure*}


DCs at the center of the Brillouin zone correspond to zero-index behavior when the appropriate homogenization criteria are met~\cite{Chan2011,li2015chip}. We perform full-wave FDTD
analysis on our DPDC structures and show that they indeed exhibit
various zero-index characteristics. One characteristic of a zero-index
medium (ZIM) is observed in the so-called ``prism''
test~\cite{li2015chip}, where plane waves normally incident on a facet
of a zero-index prism emerge at right angle from another
facet. Alternatively, one can also simulate the complex transmission
and reflection coefficients of the zero-index medium, from which
effective constitutive parameters can be extracted~\cite{Chan2011,li2015chip}.  As shown in \Figref{fig1}(c), we perform a prism test by illuminating one
side of a 45-45-90$^\circ$ triangular region made up of DPDC unit
cells and then measuring the far-field patterns emerging out of the
diagonal (hypotenuse) facet. Note that $\theta$ is the refraction
angle between the direction of the emerging beam and the facet
normal. \Figref{fig1}(d) shows smooth Gaussian beam profiles in the
far field with the refraction angles crossing zero around the
Dirac-cone wavelengths for both TE and TM polarizations. Index
retrievals~(\figref{fig1}e) also confirm the zero-index behavior
with the effective index crossing $n = 0$ between
$1.6$--$1.7~\lambda/a$. 
It must be noted that in our structures,
zero-index behavior is only observed for normal incidence;
illumination at oblique incident angles excite modes which do not
exhibit zero-index behavior, as is the case for most Dirac-cone
zero-index media~\cite{oblique15}. 

While \figref{fig1} demonstrates a perfect DPDC, the gray-scale dielectric profile poses a significant fabrication challenge (though it may be implemented in the radio frequency regime). Here, we examine a binary regularized version better suited for experimental realization. The modified structure (\figref{fig2}a) is obtained via two stages: first, we apply threshold projection filters~\cite{wang2011} \emph{during} the optimization process to produce a binary design; then, we feed the intermediate binary design into a \emph{post}-optimization pixel-averaging routine to weed out the fine features and regularize the structure (see also the supplemental material). The associated band structure~(\figref{fig2}b) shows a DPDC albeit with a \emph{spoiled} overlap due to a small frequency gap of $\sim 1\%$ between the TE and TM DCs. \Figref{fig2}(c,d) show the corresponding prism tests and index retrieval analyses. Arguably, the optimally fine-tuned gray-scale structure shows better performance than the completely
binary version. This is due to following reasons: for the gray-scale version, the zero-index crossing is perfectly linear and virtually degenerate for both polarizations;
for the binary version, the crossings are separated by about $1\%$ and
real part of the effective index shows a constant zero value while the
imaginary part depicts a bump around the zero crossing, which
corresponds to a small bandgap near the Dirac-point frequency~\cite{li2015chip}. Nevertheless, the modified structures
clearly feature a range of wavelengths where \emph{near-zero-index}
behavior is observed for both polarizations, which make them realistic candidates for practical applications. We note that an approach to realize DPDCs and
polarization-independent zero-index behavior was
recently proposed~\cite{wang2016full}, which necessitates the use of
complex meta-crystals based on patterning an anisotropic elliptic
metamaterial. In contrast, we identify DPDCs by virtue of
unconventional geometries that can be imprinted on simple ordinary
isotropic dielectrics.

The appearance of complicated features in the DPDC geometry~(\figref{fig1}) can be attributed to numerous stringent conditions imposed upon the optimization process. As noted above, one such condition is the mode separation constraint which pushes certain extraneous modes away from the Dirac degeneracy. We find that relaxing this constraint leads to a simple DPDC structure with regular geometric features~(\figref{fig3}a) although the proximity of an unintentional TM quadrupole mode in the band structure engenders an anti-crossing (aka mode mixing~\cite{joannopoulos2011photonic}) off the $\Gamma$ point near the Dirac frequency~(\figref{fig3}b). Nevertheless, FDTD analyses of \figref{fig3}(a) clearly show the near-zero-index behavior~(\figref{fig3}c,d) for both polarizations. Note that the simple ``four-hole'' structure (\figref{fig3}a) has a relative permittivity $\epsilon=3.3$, making it suitable for fabrication with common polymer materials.


{\it Dirac Cones at the $\mathbf{K}$ point.---} To demonstrate the
versatility of our approach, we proceed to design DPDCs based on a hexagonal
lattice with symmetry properties distinct from those found on a square
lattice. In particular, we focus on the $\mathbf{K}$ point of the
Brillouin zone, where two dipolar eigenmodes that transform according
to the $E$ irreducible representations of $C_{3v}$ point group form a
\emph{deterministic} DC, i.e, a DC that arises as a consequence of the symmetry of the lattice~\cite{Sakoda12,mei2012first}. We show that we can overlap two such
DCs, one with TM polarization and the other with TE polarization,
thus restoring electromagnetic duality~\cite{khanikaev2013photonic} in
the vicinity of the four-fold degenerate Dirac point. Specifically, we
employ the LDOS TO formulation to design degenerate TM and TE dipolar
modes while imposing $C_{3v}$ symmetry via suitable transformations
which ensure the concurrence of the corresponding degenerate partner
for each polarization, leading to DPDCs. 

\begin{figure*}[ht!]
\centering
\includegraphics[width=0.8\textwidth]{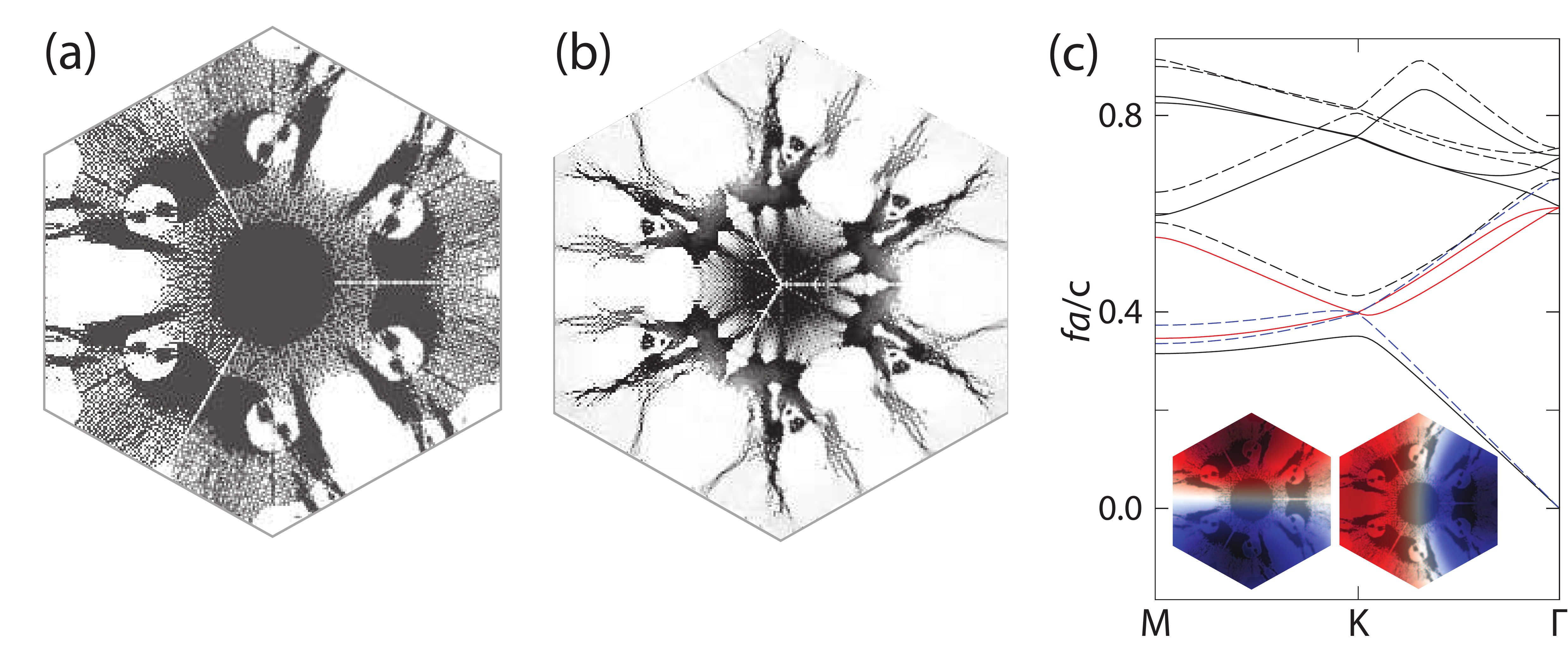}
\caption{Detailed image of (a) low-index ($\epsilon_{\rm r}=3.3$) (b) high-index ($\epsilon_{\rm r}=9$) topology-optimized hexagonal unit cell. (c) Band structure of the low-index design exhibiting overlaid TE (dashed line) and TM (solid lines) Dirac bands (red and blue). The degenerate modes (insets) transform according to the $E$ irreducible representation of $C_{3v}$ group.~\label{fig:c3v}}
\end{figure*}

\Figref{c3v} shows complex geometries discovered by TO and the
corresponding band structure with overlaid TE and TM DCs at
$\mathbf{K}$ point. The gap between the two Dirac points is as small
as $<0.1\%$, only limited by numerical discretization errors. 
To our
knowledge, this structure is the first proof-of-principle 2D design, based on
ordinary isotropic dielectric materials, that hosts overlaid TE/TM DCs at a
non-$\mathbf{\Gamma}$ point of a PhC. Moreover, this structure stands
in contrast to more sophisticated recent designs using 2D
metacrystals~\cite{khanikaev2013photonic} or 3D hexagonal
PhC~\cite{pti3d16}. Since DPDCs at the $\mathbf{K}$ point of a
hexagonal lattice are important precursors to non-trivial topological
states~\cite{lu2014topological,khanikaev2013photonic,pti3d16}, our
method suggests an alternative precursor from which one may realize a
so-called photonic topological insulator (PTI). Since our focus here is realizing DPDCs, we will not pursue making a PTI
here. However, it is worth mentioning that there are well-known
techniques to introduce non-trivial topological bandgaps into DPDCs
based on suitable bi-anisotropic perturbations, such as by introducing
off-axis propagation $(k_z \neq 0)$, by systematic reduction of mirror
symmetry, or by modifications that mix TE and TM polarizations while
preserving the pseudospin
distinction~\cite{khanikaev2013photonic}. Although the TO-discovered
geometry might be quite challenging to fabricate due to the existence
of pixel-thin hairy features, we note that these features do not
indicate a fundamental limitation of our technique but are an artifact
of underlying image-transformation steps which impose undue
constraints on the optimization process. In the supplement, we discuss
such drawbacks as well as possible ways to mitigate them.

{\it Conclusion and remarks.---} While some of the optimized designs we
have presented might prove challenging, though not impossible, to
fabricate at visible or near-infrared frequencies, they can be readily
realized at mid- to far-IR as well as microwave frequencies via
existing technologies such as computerized machining, 3D printing,
laser cutting, additive manufacturing, or two-photon
lithography, self-assembly of DNA molecules~\cite{Borisov98, Lewis15, Vukusic16,Rothemund2006}. Furthermore, thin
isolated features which typically beset topology-optimized designs can
be removed by a variety of advanced regularization and averaging
techniques~\cite{Jensen11}. The appearance of such features indicates
an optimization process that is severely constrained by the design
requirements of realizing TE and TM modes with the same modal profile
at the same frequency. The fundamental issue underlying such a design
is that in a generic structured isotropic 2D medium, TM bands tend to
be at lower frequencies than TE bands, breaking the so-called
electromagnetic duality. While we have shown that our TO formulation
is capable of restoring this duality and finding DPDCs on a 2D
lattice, this comes at the expense of a highly irregular structure
which needs to be fine-tuned with thin sensitive features. In
contrast, we surmise that three-dimensional platforms will offer even
greater flexibility. For example, it is known that TM modes tend to
experience effectively different index of refraction relative to TE
modes in 3D PhC slabs, e.g. depending on whether the PhC geometry
consists of holes or pillars~\cite{joannopoulos2011photonic}. In
future work, we will consider optimization in full 3D multi-layered geometries,
which we expect will open up even more exciting opportunities for new
structural designs in the fields of metasurfaces, metamaterials and
topological photonics.

{\it Acknowledgements.---} The authors thank Philip Camayd-Mu\~{n}oz and Orad Reshef for discussions. This work was partially supported by the Air
Force Office of Scientific Research under Contract
No. FA9550-14-1-0389, by the National Science
Foundation under Grants No. DMR-1454836 and DMR-1360889, and by the
Princeton Center for Complex Materials, a MRSEC supported
by NSF Grant No. DMR-1420541. Z. Lin is supported
by the National Science Foundation Graduate Research
Fellowship Program under Grant No. DGE1144152.

\bibliographystyle{unsrt}
\bibliography{zim,ep3}
\end{document}